\documentclass[intlimits,twoside,a4paper]{article}

\usepackage[cp1251]{inputenc}
\usepackage[eqsecnum]{cmpj3}

\usepackage{bm}

\issue{2020}{23}{4}{43714}
\doinumber{10.5488/CMP.23.43714}

\title[Electron-hole asymmetry in systems with orbital degeneracy]
{Electron-hole asymmetry in electron systems with orbital degeneracy and correlated hopping}
\author[Yu. Skorenkyy, O. Kramar, Yu. Dovhopyaty]{Yu. Skorenkyy, O. Kramar, Yu. Dovhopyaty}
\address{Ternopil Ivan Puluj National Technical University,\\56 Ruska St., 46001  Ternopil, Ukraine}

\date{Received June 15, 2020, in final form September 15, 2020}

\begin{document}

\maketitle

\begin{abstract}
Microscopic models of electronic subsystems with orbital degeneracy of energy states and non-diagonal matrix elements of electron interactions (correlated hopping) are considered within the configuration-operator approach. Equations for arbitrary-temperature numerical calculation of the doublon concentration for the integer band filling $n=1$ at different forms of the model density of states are derived. The energy spectra obtained within the Green function method are discussed with special emphasis on the role of correlated hopping in transition from itinerant to localized behavior observed in vanadium Magneli phases V$_n$O$_{2n-1}$, transition-metal dichalcogenides NiS$_{2-x}$Se$_x$, fulleride A$_n$C$_{60}$ systems. 

%\pacs 71.27.+a, 71.10.Fd, 71.30.+h
\end{abstract}

\section{Introduction}

The idea of interaction-driven electron localization in a paramagnet due to Coulomb forces (the Mott transition~\cite{mott97, geb97}) and attempts to explain ferromagnetism of transition metals have led to the elaboration of extremely simple and insightful models of quantum statistics, namely Anderson model~\cite{and61} and Hubbard model~\cite{hubb63}. The Hubbard model, describing a single non-degenerate band of electrons with the on-site Coulomb interaction, provided the mechanisms of the metal-to-insulator transition (MIT) and magnetic orderings. The model Hamiltonian contains just two energy parameters: the matrix element $t$ being the hopping integral of an electron to a neighboring site and the parameter $U$ of the intra-atomic Coulomb repulsion of two electrons on the same site. This model is used for a description of a variety of phenomena and is intensively studied~\cite{tas98,irkh07,dutt15}.
Albeit apparent simplicity, the Hubbard model does not allow for exact solution except for the very special cases of $d=1$ (one spatial dimension~\cite{lieb68}) or $d\to\infty$ (infinite spatial dimensionality~\cite{metz89}). The limit $d\to\infty$ seems to explain physics of the real $3D-$materials quite well but the wide variety of compounds in which strong electron correlations govern localization phenomena requires more than just two-parameter model to account for all of the distinctive features. To describe the peculiarities of  real materials, the matrix elements of electron interactions neglected in the original form of the Hubbard model as well as orbital degeneracy of atomic energy levels~\cite{gunn96,pav18} are to be taken into account.  

Classic materials to manifest the Mott transition are a family of vanadium Magneli phases, which have recently seen renewed attention~\cite{mog20}. In these compounds not only the metal-to-insulator transition at increasing temperatures is observed but also doping of cation subsystem or application of the external pressure determine electrical properties. The transitions in VO$_2$ and V$_2$O$_3$ can be explained within the framework of non-degenerated Hubbard model~\cite{mott97,Roz19}.  
On the contrary, Mott-Hubbard physics of transition metal dichalcogenides~\cite{wils85}, which may find new applications in next generation nanoelectronics~\cite{hu18}, is the best described when the double orbital degeneracy of $e_g$-electrons is explicitly incorporated into the Hamiltonian~\cite{prb00}.
Another class of materials with inherent orbital degeneracy and promising physical properties are fullerides~\cite{hol92}, solid-state systems with doped fullerene molecules C$_{60}$. Unusual metal-insulator transition is observed in fullerides A$_x$C$_{60}$ doped with alkali metals A. Among these compounds, A$_3$C$_{60}$ is metallic, and the other phases AC$_{60}$, A$_2$C$_{60}$ and A$_4$C$_{60}$ are insulators~\cite{poir93} while the band structure calculations predict a purely metallic behavior~\cite{sath92}. The orbital degeneracy of energy levels is responsible for the noted disagreement of theory and experiment. The Gutzwiller variational approach~\cite{lu94} predicts MIT for all integer band fillings $n=1, 2, 3, 4, 5$. Experimentally measured electrical resistivity of polycrystals C$_{60}$~\cite{regu91} monotonously changes with temperature and the energy gap has a monotonous dependence on the external pressure.

A breakthrough in solving the puzzle of strong correlation happened due to the development of the dynamical mean-field theory (DMFT) in the limit of infinite dimensionality~\cite{metz89,DMFT96}. It was quickly realized that DMFT is not applicable to the system with the hopping amplitudes dependent on the site occupancies~\cite{schi99}. Studies of Falicov-Kimball model~\cite{fal69,free03}, a partial case of the Hubbard model with two sorts of particles, one localized and the other itinerant, shed light on the peculiarities of the systems with correlated hopping of electrons (dependency of hopping amplitudes on lattice site occupancies). Within DMFT for Falicov-Kimball model, an exact two-pole structure was obtained for Green function, while for a general case of Hubbard model, a finite-temperature perturbation theory with electron hopping term as a perturbation parameter was built~\cite{shv00}.On this basis, a rigorous approach for description of correlated hopping within the Falicov-Kimball model was built, thermodynamics of the model and temperature-induced MIT were studied~\cite{shv03,shv03pss}. Within the elaborated scheme, considerable progress was achieved in theoretical description of inelastic Raman and X-ray scattering in the vicinity of MIT~\cite{shv04,pakh12} and in charge-density wave phase~\cite{mat09,mat16}. The model with correlated hopping was shown to manifest a peculiar optical conductivity~\cite{dob18} and thermoelectric properties~\cite{shv14}--\cite{dob20} due to the correlated hopping of electrons. 

The present study is devoted to investigation of a system close to Mott-Hubbard transition within the  models taking into account the orbital degeneracy of energy levels, strong Coulomb interaction and correlated hopping of electrons.
The structure of this paper is as 
follows. Section \ref{sec:Ham} is devoted to the model formulation and model 
Hamiltonians for non-degenerated, doubly and triply degenerated models. The general representations of the Hamiltonians in the electron and Hubbard operators is given. In section~\ref{sec:results} the concentrations of empty sites (holes) and doubly occupied sites close to metal-insulator transition at the electron concentration $n=1$ for models with non-degenerated, doubly- and triply-degenerated energy levels are calculated. For calculation of quasiparticle energy spectra, the nonperturbative approach, generalized mean-field approximation~\cite{did98,cmp08} is used. Section \ref{sec:concl} presents our conclusions on the role of correlated hopping in models with orbital degeneracy.

\section{The Hamiltonians of electronic subsystem with orbitally degenerated energy levels}\label{sec:Ham}

Within the second quantization formalism, the Hamiltonian of interacting electron systems can be written~\cite{fett71} as
\begin{eqnarray}
\label{gen_Ham}
H= -\mu \sum_{i\lambda\sigma} a_{i\lambda\sigma}^+ a_{i\lambda\sigma}
+{\sum_{ij\lambda\sigma}}'t_{ij}a_{i\lambda\sigma}^+ a_{j\lambda\sigma}+
\frac{1}{2}{\sum_{ijkl}}{\sum_{\alpha\beta\gamma\delta}}{\sum_{\sigma\sigma'}}J^{\alpha\beta\gamma\delta}_{ijkl}
a_{i\alpha\sigma}^+a_{j\beta\sigma'}^+ a_{l\delta\sigma'}a_{k\gamma\sigma}\,,
\end{eqnarray}
where $\mu$ stands for chemical potential, $a_{i\lambda\sigma}^{+}$, $a_{i\lambda\sigma}$ are operators of spin-$\sigma$ electron creation and annihilation in orbital state $\lambda$ on lattice site $i$, respectively, the second sum with matrix element
\begin{eqnarray}
t_{ij}=\int{\rd^3} r{\phi}_{\lambda}^{*}({\bf r}-{\bf R}_{i})
\left[ -\frac{\hbar^2}{2m}\Delta +V^{\text{ion}}({\bf r})\right]
\phi_{\lambda} ({\bf r}-{\bf R}_{j})
\end{eqnarray}
describes translations (hopping) of electrons in the crystal field $V^{\text{ion}}(\bf{r})$. The third sum in equation (\ref{gen_Ham}) is the general expression for pair electron interactions described by matrix elements
\begin{eqnarray}
J^{\alpha\beta\gamma\delta}_{ijkl}&=&\int{\int{{\phi}_\alpha^{*}({\bf r}-{\bf R}_{i}){\phi}_\beta({\bf r}-{\bf R}_{j})}}
\frac{\re^2}{|r-r'|}{\phi}_\delta^{*}({\bf r'}-{\bf R}_{l}){\phi}_\gamma({\bf r'}-{\bf R}_{k})\rd r \rd r'.
\end{eqnarray}
In the above formulae,  indices $\alpha$, $\beta$, $\gamma$, $\delta$, $\lambda$ denote orbital states, ${\phi}_{\lambda i}$ is wave-function in Wannier (site) representation, other notations are standard. 

Hamiltonian~(\ref{gen_Ham}) is hard to treat mathematically due to its non-diagonal form in both Wannier- and wave-vector representation. The model is greatly simplified if all matrix elements are neglected except on-site Coulomb correlation (the Hubbard parameter $U$):
\begin{eqnarray}
U=\int{\int{|{\phi}_\lambda^{*}({\bf r}-{\bf R}_{i})|^2 \frac{\re^2}{|r-r'|}|\phi_\lambda ({\bf r'}-{\bf R}_{i})|^2\rd r \rd r'}}.
\end{eqnarray}
In this case, however, we fail to describe the influence of lattice site occupation on the electron hopping (correlated hopping of electrons) which removes the electron-hole symmetry for real correlated electron systems~\cite{prb00} as well as plays an important role in the onset of itinerant ferromagnetism~\cite{amad_hir,volh99,kollar,fark,prb01}. Besides, for a system with orbital degeneracy of energy states, the on-site exchange integral (Hund's rule coupling)
\begin{eqnarray}
J_H&=&\int{\int{\phi}_\lambda^{*}}({\bf r}-{\bf R}_{i})\phi_{\lambda^{'}}({\bf r}-{\bf R}_{i}){\re^{2}\over |{r}-{r}^{'}|}
\phi^{*}{_{\lambda^{'}}}({\bf r}^{'}-{\bf R}_{i})\phi_\lambda({\bf r}^{'}-{\bf R}_{i})\rd{\bf r}\rd{\bf r}^{'},
\end{eqnarray}
which makes the configurations with parallel spins energetically more favorable, should be taken into account. 

The Hamiltonian of the generalized Hubbard model with correlated hopping of electrons then reads as
\begin{eqnarray}
H_{s}&=& -\mu \sum_{i\sigma} n_{i\sigma}+ U\sum_{i}n_{i\uparrow}n_{i\downarrow}+
{\sum_{ij\sigma}}'t_{ij}(n)a_{i\sigma}^+ a_{j\sigma}+
{\sum_{ij\lambda\sigma}}'T^{'}_{ij}\left(a_{i\sigma}^+ a_{j\sigma}n_{i\bar\sigma}+h.c. \right),
\end{eqnarray}
where $n_{i\sigma}=a_{i\sigma}^{+}a_{i\sigma}$ is the site occupancy operator, hopping integrals $t_{ij}(n)$, $T^{'}_{ij}$ taking into account two types of correlated hopping of electrons~\cite{did98} in non-degenerate model are introduced. Such a generalized model is not invariant under particle-hole transformation $a_{i\sigma}^+ \to a_{i\sigma}$, in distinction from the Hubbard model~\cite{miel15}.  

The Hamiltonian of the orbitally degenerated electronic system with strong electron correlation and correlated hopping of electrons can be represented in the form
\begin{eqnarray}
&H_{\text{deg}}&=H_{\text{loc}}+H_{\text{tr}}\,,
\\
&H_{\text{loc}}&= -\mu \sum_{i\lambda\sigma} n_{i\lambda\sigma}+
U\sum_{i\lambda}n_{i\lambda\uparrow}n_{i\lambda\downarrow}+
\frac{U'}{2}\sum_{i\lambda\sigma}n_{i\lambda\sigma}n_{i\lambda'\bar\sigma}+
\frac{U'-J_H}{2}\sum_{i\lambda\lambda'\sigma}n_{i\lambda\sigma}n_{i\lambda'\sigma}\,,
\nonumber
\\
&H_{\text{tr}}&={\sum_{ij\lambda\sigma}}'t_{ij}(n)a_{i\lambda\sigma}^+ a_{j\lambda\sigma}+
{\sum_{ij\lambda\sigma}}'t^{'}_{ij}\left(a_{i\lambda\sigma}^+ a_{j\lambda\sigma}n_{i\bar\lambda}+h.c. \right)+
{\sum_{ij\lambda\sigma}}'t^{''}_{ij}\left(a_{i\lambda\sigma}^+ a_{j\lambda\sigma}n_{i\lambda\bar\sigma}+h.c. \right),
\nonumber
\end{eqnarray}
where $n_{i\lambda\sigma}=a_{i\lambda\sigma}^{+}a_{i\lambda\sigma}$, $U'=U-2J_H$ and hopping integrals $t_{ij}(n)$, $t^{'}_{ij}, t^{''}_{ij}$ are introduced taking into account different types of correlated hopping of electrons~\cite{prb00}.
In a model of doubly degenerate band, every site can be in one of 16 configurations~\cite{prb00} while for triple degeneracy, inherent to fullerides, the number of possible configurations rises to 64~\cite{upj12}.

In fullerenes, which have triply degenerated lowest unoccupied molecular orbital of $t_{1u}$ symmetry, the competition between intra-site Coulomb interaction and electron hoppings defines the electrical properties, with insulator or metallic behavior realized~\cite{gunn96}.
Fullerides (fullerene crystals) are semiconductors with an energy gap of $1.2-1.9$~eV~\cite{sait91,achi91}. Intra-atomic repulsion energy estimates~\cite{hett91} give 
$2.7$ eV. Calculation with a screening effect taken into account gives $U=2.7$~eV~\cite{ped92,antr92}. The energy cost of electron configurations with two spins aligned in parallel on different orbitals is considerably less than for anti-parallel alignment due to Hund's rule coupling.

To consider the case of strong electron correlation, it is reasonable to pass from electron operator to Hubbard operators $X^{pl}$ of site transition from state $| l \rangle$ to state $| p \rangle$ with anticommutation relations $\{X_{i}^{pl};X_{j}^{kt}\}=\delta_{ij}(\delta_{lk}X_{i}^{pt}+\delta_{pt}X_{i}^{kl})$, and normalizing condition $ \sum \limits_{i}X_i^p=1$, for number operators $X_i^p=X_{i}^{pl}X_{i}^{lp}$ of $|p\rangle$--state on site $i$.

In the configuration-operator representation, the model Hamiltonian takes the form 
\begin{eqnarray}
\label{eff_ham_fm}
H=H_{\text{loc}}+\sum_{\lambda=\alpha,\beta,\gamma}\left(H_b^{(\lambda)}+H_h^{(\lambda)}\right),
\end{eqnarray}
where the interaction part $H_{\text{loc}}$ is diagonal, the processes forming Hubbard subbands described by $H_b^{(\lambda)}$ and the hybridization of these subbands $H_h^{(\lambda)}$ are clearly separated in the translational part of the Hamiltonian, and  which constitute a distinctive feature of configuration-operator representation. Different hopping integrals correspond to transitions in (or between) the different subbands. 

In the partial case of band filling $n=1$, strong Coulomb correlation and strong Hund's coupling (parameter $U-3J_H$ is much greater than the bandwidth), the states with three and more electrons on the same site are excluded. Then, the influence of correlated hopping can be described by three different hopping integrals. The  bare band hopping integral $t_{ij}$ is renormalized to take into account the band narrowing caused by concentration dependent correlated hopping as $t_{ij}(n)=t_{ij}(1-\tau_1n)$. This hopping integral characterizes the lower Hubbard subband. Parameter $\tau_1$ is usually neglected, but it is of principal importance for a consistent description of correlation effects in narrow band systems. The hopping integral for the upper Hubbard subband is $\tilde{t}_{ij}(n)=t_{ij}(n)+2t'_{ij}=(1-\tau_1 n-2\tau_2)t_{ij}$  and $\bar{t}_{ij}(n)=t_{ij}(n)+t'_{ij}=(1-\tau_1 n-\tau_2)t_{ij}$ describes a hybridization of lower and upper Hubbard subbands, correlated hopping parameter $\tau_2$ describes the effect of occupancy of the sites involved in the electron transfer and $\tau_1$ describes that of the nearest-neighbor sites.  

\section{Results and discussion}\label{sec:results}

Green functions technique allows us to calculate the energy spectrum of the model in the case of electron concentration $n=1$. One can rewrite the single-particle Green function $\langle\langle a_{i\lambda \sigma} | a_{j\lambda \sigma}^{+}\rangle\rangle$ on the basis of the relation between electronic operators and Hubbard's $X$-operators describing electron hoppings between holes, single-electron configurations and Hund's doublons. Probabilities of the processes involving another type of doubly occupied states, states with three or more electrons tend to zero in the considered limit of strong intra-atomic correlation.

Substantial extension of the Hilbert space for the cases with orbital degeneracy makes the direct application of standard methods quite complicated~\cite{pav17}. One may considerably simplify the calculation of quasiparticle energy spectrum  by projecting the terms arising from commutation with $H_{\text{tr}}$ in the Green function equation of motions onto a set of basic operators which describe processes responsible for Hubbard subbands formation. The configuration representation of Hubbard operators is especially well-suited for such procedures and makes physical nature of the metal-to-insulator transition and the influence of correlated hopping relatively transparent. In the limit of strong Coulomb correlation and strong Hund's coupling, for the doubly degenerated band, such set of operators is $X^{0,\lambda\sigma}_i$, $X^{\lambda\sigma,2}_i$, and for triply degenerated model it consists of $X^{0,\lambda\sigma}_i$ and $Y^{\lambda\sigma,2}_i$ where $Y$-operator is introduced to take into account that for the triply degenerate orbitals two equivalent states are involved, for example, $Y^{\alpha\uparrow ,2}_i=X^{0\uparrow 0,\uparrow\uparrow 0}_i+X^{00\uparrow,\uparrow 0\uparrow}_i$. 

To obtain a closed system of equations for Green functions $\langle\langle X_{p}^{0,\alpha\uparrow}|X_{p'}^{\alpha\uparrow0} \rangle\rangle$  and $\langle\langle Y_{p}|X_{p'}^{\alpha\uparrow,0} \rangle\rangle$, we use the projection procedure~\cite{did98}:
\begin{eqnarray}
\label{project}
[X_{p}^{0,\alpha\uparrow};\sum_\lambda{H_b^{(\lambda)}}]&=&\sum_{i}\varepsilon_{pi}^bX_{i}^{0,\uparrow};
\qquad
[Y_{p};\sum_\lambda{H_b^{(\lambda)}}]=\sum_{i}\tilde \varepsilon_{pi}^bY_i;
\\
\nonumber
[X_{p}^{0,\alpha\uparrow};\sum_\lambda{H_h^{(\lambda)}}]&=&\sum_{i}\varepsilon_{pi}^hY_i;
\qquad \quad
[Y_{p};\sum_\lambda{H_h^{(\lambda)}}]=\sum_{i}\tilde \varepsilon_{pi}^hX_{i}^{0,\uparrow}.
\nonumber
\end{eqnarray}
This procedure realizes the generalized mean-field approximation which allows one to easily break off the sequence of Green function equations and results in two-pole quasiparticle Green function, with the spectrum describing two Hubbard subbands with renormalized widths and hybridization. To obtain the single-electron Green function, however, one should solve two systems of equations (for $\langle\langle X_{p}^{0,\alpha\uparrow}|X_{p'}^{\alpha\uparrow0} \rangle\rangle$, $\langle\langle Y_{p}|X_{p'}^{\alpha\uparrow,0} \rangle\rangle$ and another for $\langle\langle X_{p}^{0,\alpha\uparrow}|Y^+_{p} \rangle\rangle$, $\langle\langle Y_{p}|Y^+_{p}\rangle\rangle$), neglecting the irreducible parts. Through the non-operator coefficients $\varepsilon^b({\bf k}),\tilde\varepsilon^b({\bf k}),\varepsilon^h({\bf k}),\tilde\varepsilon^h({\bf k})$, the spectrum becomes dependent on the mean numbers of sites in particular electron configurations (for example, Hund doublons in the doubly orbitally degenerated case) and thus dependent on temperature. 
   
Solving analytically the system of equations for $\langle\langle X_{p}^{0,\alpha\uparrow}|X_{p'}^{\alpha\uparrow0} \rangle\rangle$  and $\langle\langle Y_{p}|X_{p'}^{\alpha\uparrow,0} \rangle\rangle$ after Fourier transformation, we obtain the quasi-particle energy spectrum
\begin{eqnarray}
\label{spec}
E_{1,2}({\bf k})=&-&\mu+\frac{\Delta}{2}+\frac{\varepsilon^b({\bf k})+\tilde\varepsilon^b({\bf k})}{2}
\nonumber
\\
&\mp& \frac{1}{2}\sqrt{(\Delta-\varepsilon^b({\bf k})+\tilde\varepsilon^b({\bf k}))^2+4\varepsilon^h({\bf k})\tilde\varepsilon^h({\bf k})}
\end{eqnarray}
with $\Delta=U$ corresponding to the non-degenerated Hubbard model and $\Delta=U-3J_H$ being the energy cost of creating a Hund's doublon --- a pair of electrons with parallel spins in different orbitals of the same lattice site~\cite{prb00,pav18} for the models with orbital degeneracy. In a more robust approach of work~\cite{shv00}, the two-pole structure of the Green function was found to realize only for Falicov-Kimball model while in general four state, subspaces (electron, hole and resonant-valence bond) were considered. In terminology of papers~\cite{shv00,shv01}, we study here only the low-energy sector of the model to compare the effects produced by the degeneracy of the energy levels and the correlated hopping, both removing the particle-hole symmetry. Concentrations of doublons appear to be an important parameter in this approach that allows one to naturally introduce a dependence of the energy spectrum~(\ref{spec}) on temperature or doping. In the degenerate model, in distinction from the original Hubbard model, a doublon is not just charge excitation but may carry a spin as well. For sufficiently high temperatures, in the absence of any type of magnetic order, we denote the concentration of empty lattice sites (holes) by $c$, concentration of sites occupied by a single electron of any spin projection $\sigma$ in any orbital state $\lambda$ by $s$, Hund's doublons  concentration by $d$ and neglect all other configurations. For the case of strong Hund's coupling, the high energy doublon configurations (spin-singlet Hubbard doublons and spin-triplet non-Hund doublons) are excluded. We make use of the completeness condition for the $X$-operator set to obtain constraints $c+2s+d=1$ for a non-degenerate model, $c+4s+6d=1$ for a model with double degeneracy, $c+6s+6d=1$ for a model with triple degeneracy at strong Coulomb correlation and at strong Hund's coupling.

The non-operator coefficients $\varepsilon^b({\bf k}),\tilde\varepsilon^b({\bf k}),\varepsilon^h({\bf k}),\tilde\varepsilon^h({\bf k})$ can be obtained by a procedure described in detail in paper~\cite{cmp08} in the paramagnetic case at $n=1$ for the doubly degenerated model as 
\begin{eqnarray}
\label{coef2}
&&\varepsilon^b=\left((1-\tau_1)\frac{1+4d+12(1-4d)^2}{4(1+4d)}+(1-\tau_1-2\tau_2)\frac{8d^2}{1+4d}\right)\varepsilon;
\\
&&\varepsilon^h=(1-\tau_1-\tau_2)d(3+4d)\varepsilon,
\\
&&\tilde\varepsilon^b=\left(-8d^2(1-\tau_1)+\frac{(1-\tau_1-2\tau_2)}{8}(1-16d+32d^2)\right)\varepsilon,
\\
&&\tilde\varepsilon^h=(1-\tau_1-\tau_2)\frac{d(3+4d)}{1+4d}\varepsilon
\end{eqnarray}
and for the triply degenerated paramagnet at $n=1$ as
\begin{eqnarray}
\label{coef3}
&&\varepsilon^b=\left((1-\tau_1)\frac{216d^2-12d+1}{24d+1}+(1-\tau_1-2\tau_2)\frac{72d^2}{24d+1}\right)\varepsilon;
\\
&&\varepsilon^h=(1-\tau_1-\tau_2)\frac{8d-12d^2}{1-6d}\varepsilon,
\\
&&\tilde\varepsilon^b=\left(-(1-\tau_1)\frac{36d^2}{1-6d}+(1-\tau_1-2\tau_2)\frac{1-16d+84d^2}{2(1-6d)}\right)\varepsilon,
\\
&&\tilde\varepsilon^h=(1-\tau_1-\tau_2)\frac{24d+1-216d^2}{3(24d+1)}\varepsilon.
\end{eqnarray}
Corresponding expressions for non-degenerated  model~\cite{cmp08} are
\begin{eqnarray}
\label{coef1}
&&\varepsilon^b=\left[(1-\tau_1)(1-2d+2d^2)-2(1-\tau_1-2\tau_2)d^2\right]\varepsilon;
\\
&&\tilde\varepsilon^b=\left[(1-\tau_1-2\tau_2)(1-2d+2d^2)-2(1-\tau_1)d^2\right]\varepsilon;
\\
&&\varepsilon^h=\tilde\varepsilon^h=-2(1-\tau_1-\tau_2)d\varepsilon.
\end{eqnarray}
One can see that in degenerated models there is no symmetry between expressions describing hole and doublon sectors.

For the generalized Hubbard model with correlated hopping~\cite{cmp08}, the doublon concentration is given by
\begin{eqnarray}
\label{doubl}
d=C\int^{w}_{-w}\rho(\varepsilon){\left[\frac{A(\varepsilon)}{\exp(\frac{E_1(\varepsilon)}{\Theta}+1)}+\frac{B(\varepsilon)}{\exp(\frac{E_2(\varepsilon)}{\Theta}+1)}\right]}\rd\varepsilon,
\end{eqnarray}
where $C={1\over2}$, $\Theta=k_{\text B}T$,
\begin{eqnarray}
\nonumber
&&A({\bf k})={1\over 2}\left(1+\frac{\Delta+\tilde{\varepsilon}^b-\varepsilon^b}
{\sqrt{(\Delta-\varepsilon^b+\tilde{\varepsilon}^b)^2+4\varepsilon^h\tilde{\varepsilon}^h}}\right),
\\
&&B({\bf k})=1-A({\bf k}).
\end{eqnarray}
The equation for holes for non-degenerated model is completely identical to the above equation as $c=d$, so $\mu$ is obtained from condition $n=1$.

For the doubly degenerated model with correlated hopping, the doublon concentration is determined by the condition~(\ref{doubl}) with $C={1\over4}$ to be solved together with the condition for the chemical potential $c=2d$, which for the doubly degenerated model can be written as 
\begin{eqnarray}
8d=(1+4d)\int^{w}_{-w}{\rho(\varepsilon)\left(\frac{B(\varepsilon)}{\exp(\frac{-E_1(\varepsilon)}{\Theta}+1)}+\frac{A(\varepsilon)}{\exp(\frac{-E_2(\varepsilon)}{\Theta}+1)}\right)}d\varepsilon.
\end{eqnarray}

In analogous way we obtain equation~(\ref{doubl}) with $C={1\over12}$ for triply degenerated model with correlated hopping.
To complete a set of equations, the condition for the chemical potential $c=6d$ for the triply degenerated model is also used 
\begin{eqnarray}
36d=(1-24d)
\int^{w}_{-w}{\rho(\varepsilon)\left[\frac{B(\varepsilon)}{\exp(\frac{-E_1(\varepsilon)}{\Theta}+1)}+\frac{A(\varepsilon)}{\exp(\frac{-E_2(\varepsilon)}{\Theta}+1)}\right]}\rd \varepsilon,
\end{eqnarray}

Solving these equations numerically with appropriate model density $\rho(\varepsilon)$ of electronic states (DOS), we obtain the doublon concentration as functions of the model parameters and may apply the gap criterion for the spectra~(\ref{spec}) to study a metal-insulator transition (MIT)~\cite{geb97,wils85}. In the ground state, at the point of MIT, the polar states (holes and doublons) concentrations tend to zero. 

In distinction from the non-degenerated model, where it is the correlated hopping which destroys the particle-hole symmetry, in doubly and triply-degenerated models, there is no symmetry even at $\tau_1=\tau_2=0$ due to non-equivalence of subspaces for empty states and doublon states (see~\cite{shv03}). 

\begin{figure}[!t]
	\begin{center}
		\begin{minipage}[h]{0.48\linewidth}
			\includegraphics[width=1\linewidth]{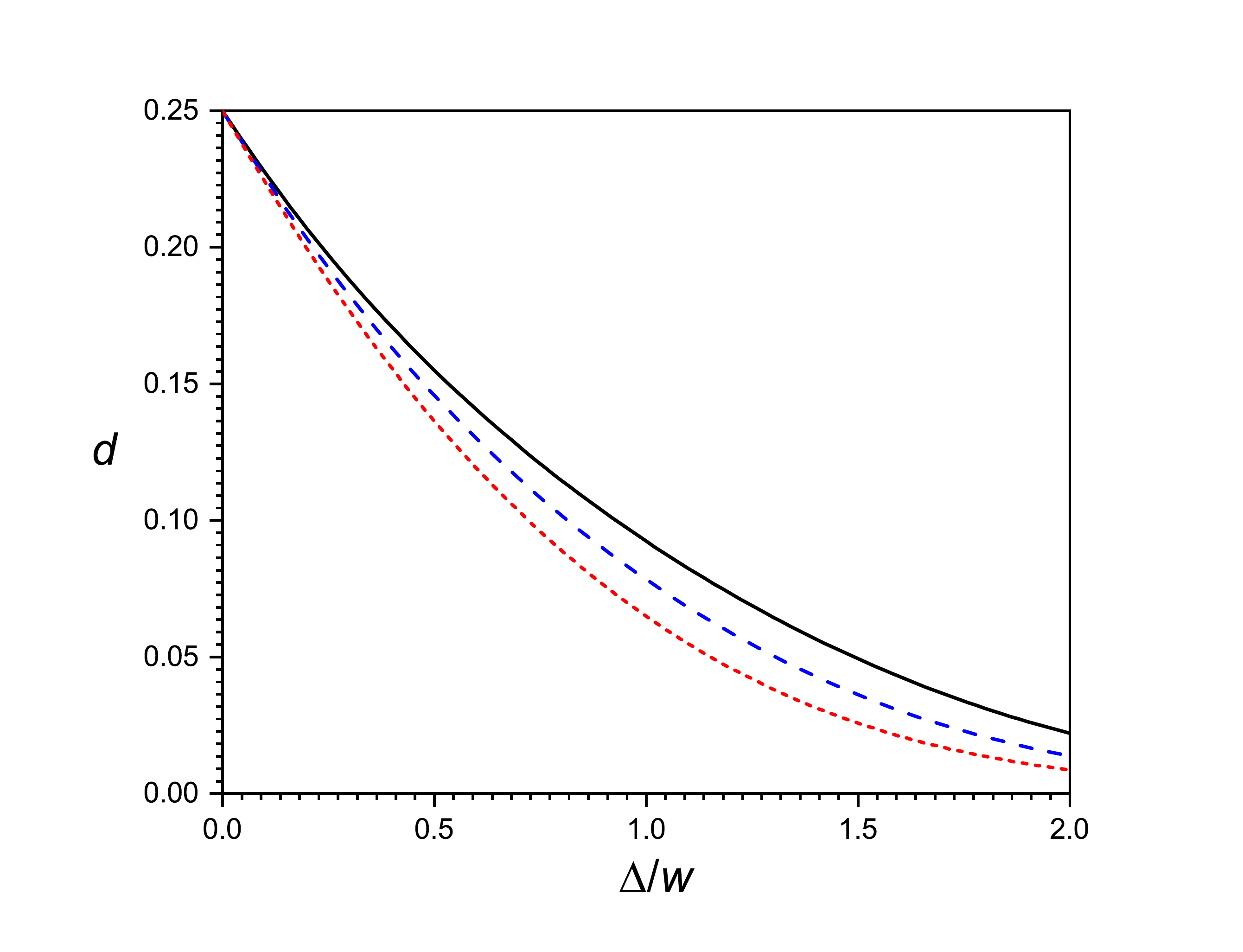}
			\caption{(Colour online) Dependence of doublon concentration on the energy parameter $\Delta /w$ for different correlated hopping values at $\Theta / w=0.2$ for non-degenerated model. Solid curve corresponds to the Hubbard model $\tau_1=\tau_2=0$, short-dashed curve corresponds to $\tau_1=\tau_2=0.1$ and long-dashed curve corresponds to $\tau_1=\tau_2=0.2$. Semi-elliptic DOS is used in calculations.}
			\label{fig1}
		\end{minipage}
		\hfill
		\begin{minipage}[h]{0.48\linewidth}
		\vspace{-0.8cm}
			\includegraphics[width=1\linewidth]{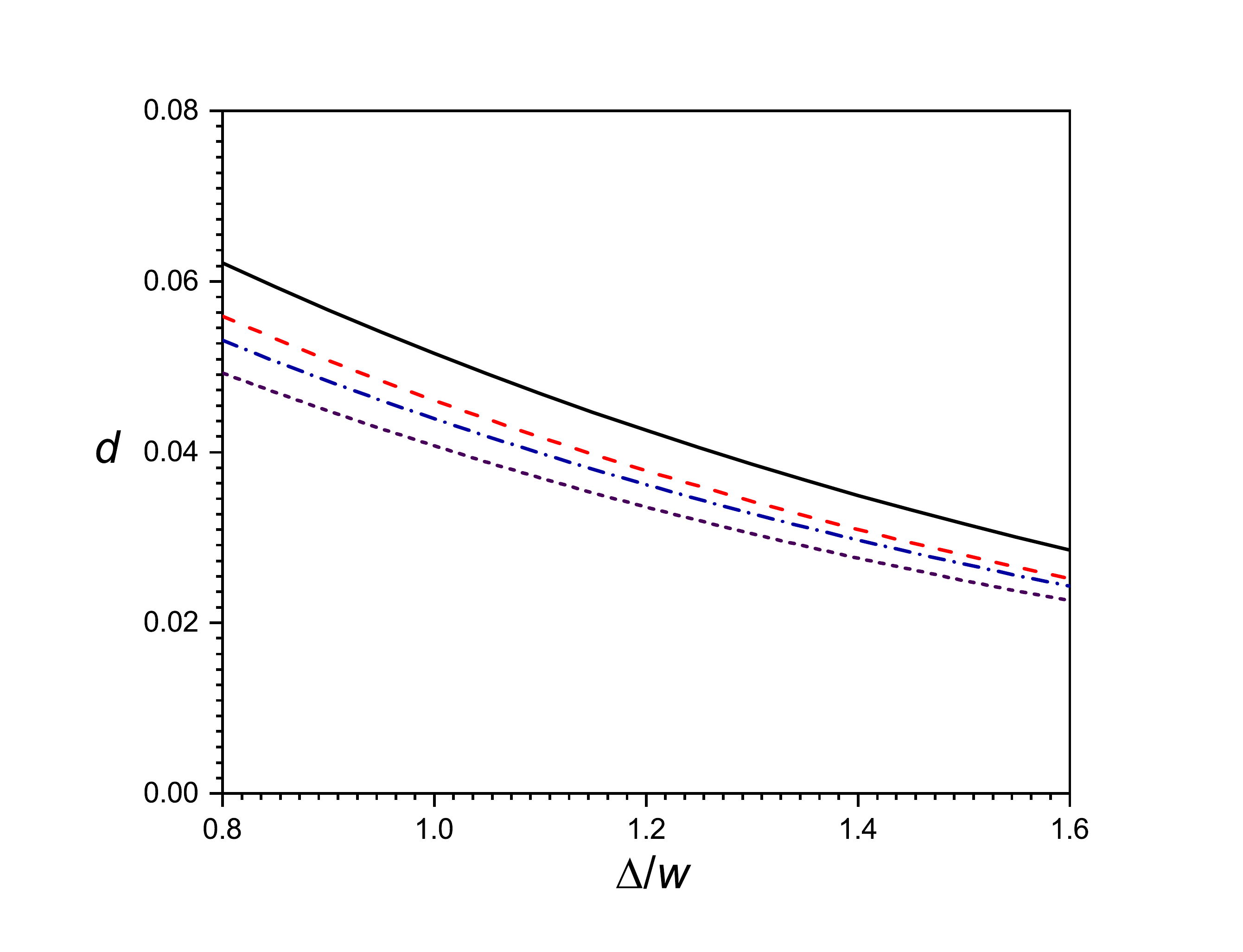}
			\caption{(Colour online) Dependence of doublon concentration in doubly orbitally degenerated model  at $\Theta / w=0.4$ on the energy parameter $\Delta /w$ for DOS with asymmetry on a band edge at different values of the asymmetry parameter $a=0, 0.3, 0.5, 0.99$ from upper to lower curve.}
			\label{fig2}
		\end{minipage}
	\end{center}
\end{figure}

Dependencies of the doublon concentration on the correlation strength, shown in figures~\ref{fig1}, \ref{fig2}, allow us to illustrate the influence of  correlated hopping and asymmetry of the DOS form.

\begin{figure}[!t]
	\begin{center}
		\begin{minipage}[h]{0.48\linewidth}
		\includegraphics[width=1\linewidth]{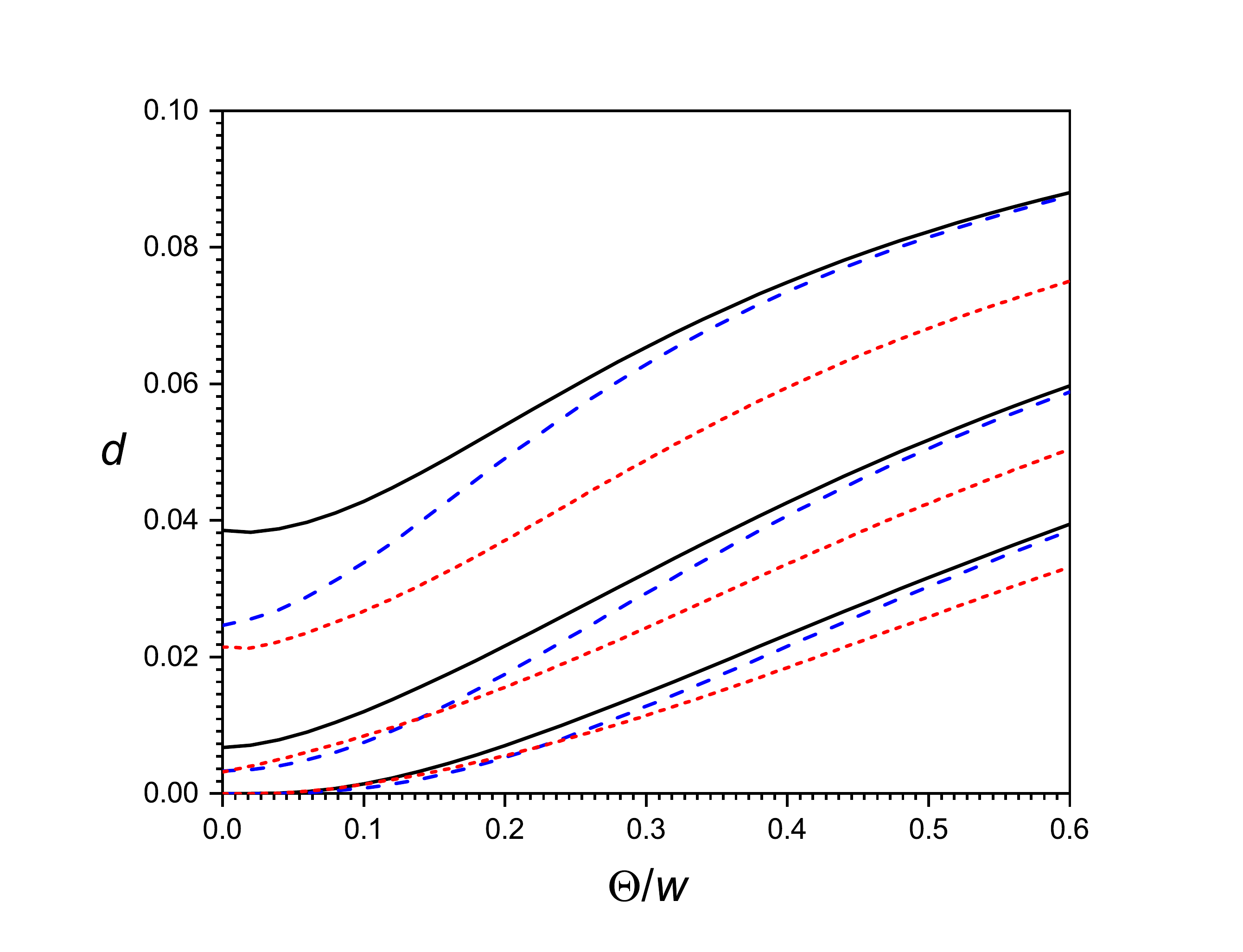}
			\caption{(Colour online) Dependence of doublon concentration in doubly orbitally degenerated model on temperature at fixed values of the energy parameter $\Delta / w=0.6$ for the upper group of curves, $\Delta / w=1.2$ for the middle group of curves and $\Delta / w=1.8$ for the lower group of curves. Solid curves correspond to the semi-elliptic DOS, long-dashed curves correspond to simple cubic lattice DOS and short-dashed curves correspond to DOS with asymmetry.}
			\label{fig3}
		\end{minipage}
		\hfill
		\begin{minipage}[h]{0.48\linewidth}
		\vspace{-0.7cm}
			\includegraphics[width=1\linewidth]{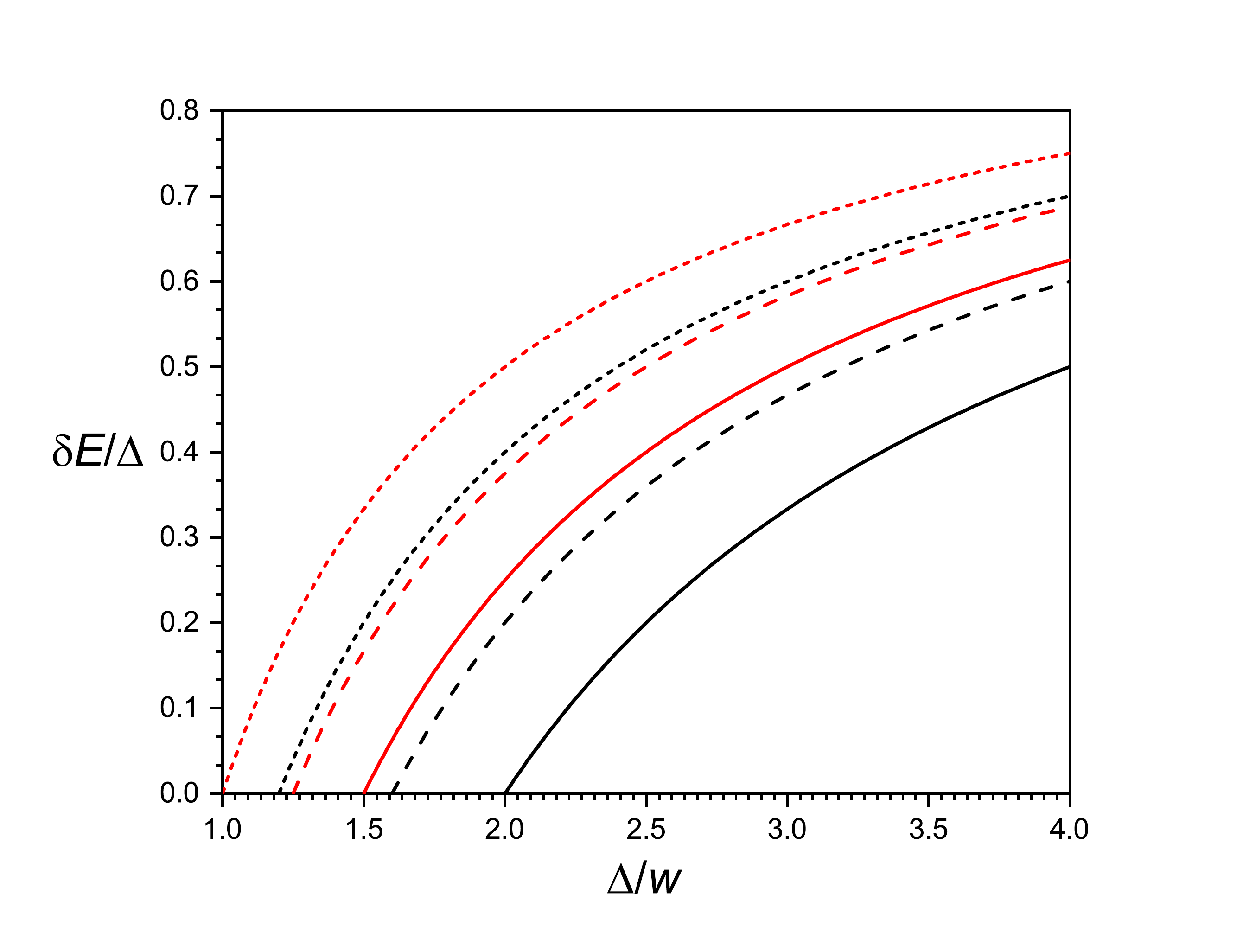}
			\caption{(Colour online) Dependence of energy gap on the interaction strength parameter for non-degenerated and triply degenerated models. In each pair, the lower curve corresponds to non-degenerated model and the upper one corresponds to the triply-degenerated model: for solid curves $\tau_1=\tau_2=0$, for long-dashed curves  $\tau_1=\tau_2=0.1$, for short-dashed curves  $\tau_1=\tau_2=0.2$.}
			\label{fig4}
		\end{minipage}
	\end{center}
\end{figure}

One can see from figures \ref{fig1}--\ref{fig3} that the correlated hopping substantially influences the doubly occupied states concentration in a system with orbital degeneracy of the energy levels. Quite naturally, an increase of the correlated hopping suppresses the conductance through electron localization, and the doublon concentration decreases, whose effect for non-degenerated model can be seen from figure~\ref{fig1}. In a way, the role of the correlated hopping does not change when the models with orbital degeneracy are considered, but those models offer more mechanisms for particle-hole asymmetry than just the correlated hopping. The DOS form can also affect the probability of doublon and hole pair creation. To study an effect of bare-band density of states asymmetry on the energy spectrum of a model with orbital degeneracy, we make use of a model DOS with asymmetry on the band edge~\cite{volh99}. If the density of states is not symmetrically distributed with respect to the subband center, the doublon concentration is smaller than for a semi-elliptical DOS (see figure~\ref{fig2}) and $\rho(\varepsilon)$ makes a more prominent effect for weaker correlations. DOS form asymmetry was shown to make more preferable conditions for ferromagnetism stabilization~\cite{volh99}. From our results we conclude that both DOS asymmetry and correlated hopping favor electron delocalization and ferromagnetism (see also~\cite{prb01} where a similar effect was observed in the case of week correlation). Figure~\ref{fig3} visualizes a temperature dependence of the doublon concentration. Three families of dependencies are shown, each corresponding to a fixed value of the correlation strength $\Delta=U-3J_H$. The dependency for DOS form which corresponds to a simple cubic lattice of transition metal compounds with double orbital degeneracy of $e_g$ band is depicted by long-dashed curves. One can see that the model DOS with tunable asymmetry parameter provides lower (at high asymmetries) and upper (for the limiting case of semi-elliptical DOS) bounds for doublon concentration with sc-lattice DOS. Qualitatively, the temperature dependencies monotonously increase in the whole temperature interval, but are flattened at the greater values of the interaction parameter. Both the occupancy of the sites involved into the hopping processes (through the correlated hopping of the first type) and the neighbor sites  (through the second type correlated hopping), have the effect of a gap in the energy spectrum and stabilization of the insulator state. 
The energy gap, however, will not open up until a relatively large increase of correlated hopping parameters takes place, not achievable in real systems. At an increase of intra-site interaction parameter over a critical value (dependent on the correlated hopping strength), the energy gap occurs and the metal-insulator transition takes place (see figure~\ref{fig4} for zero temperature).  An increase of the polar states (holes and doublons) concentration with temperature change modifies the energy spectrum, the energy gap opens up and the insulating state stabilizes. On the contrary, the application of the external pressure or doping in systems (V$_{1-x}$Cr$_x)_2$O$_3$ and NiS$_{2-x}$Se$_x$ leads to metalization~\cite{mott97} by increasing the energy subbands width and reducing $\frac{\Delta}{w}$ below the critical value. The critical values for the partial case when there is no correlated hopping are in agreement with the results of paper~\cite{cmp08} for non-degenerated Hubbard model and paper~\cite{upj12} for a triply degenerated model.

\section{Conclusions}\label{sec:concl}

Continuing the studies of non-denenerate Hubbard model generalized with taking into account the correlated hopping of electrons, quasiparticle energy spectra for doubly- and triply degenerated model are calculated to show that both the orbital degeneracy and the correlated hopping remove the particle-hole symmetry and have a strong effect on the electron localization. The energy spectra of the lower and upper subband depend on the polar states concentration and are essentially non-equivalent.

Equations are derived for an arbitrary-temperature numerical calculation of the doublon concentration for the integer band filling $n=1$ at different forms of the model density of states. 
Within the models of strongly correlated electron systems with orbitally degenerated energy levels, both the correlated hopping and the degeneracy of energy levels are crucial to describe the effects of particle-hole asymmetry. The use of Hubbard $X$-operators representation appears to be useful to structure the model Hamiltonian according to the occupancies of the state subspaces and allows one to single out the quasiparticle subbands, relevant for metal-insulator transition. The expression for the energy gap which is temperature dependent through the polar state concentration, which also depends on the DOS form, alows one to describe the phase transition under the external actions. 
Critical values of the correlation strength parameter for correlation-driven metal-insulator transition in models with the degeneracy of energy levels and correlated hopping  substantially differ from that of Hubbard model. 
Taking into account a number of microscopic parameters can be a leverage for tuning the analytical expressions to reveal the relevant microscopic mechanisms of electron localization in materials with non-equivalent subbands.

\ukrainianpart

\title{Електрон-дiркова асиметрiя в системах з орбiтальним
виродженням та корельованим переносом}
\author{Ю. Скоренький, О. Крамар, Ю. Довгоп’ятий}
\address{
 Тернопiльський нацiональний технiчний унiверситет iменi Iвана Пулюя, вул. Руська, 56, \\46001 Тернопiль, Україна 
}

\makeukrtitle

\begin{abstract}
У роботi дослiджено мiкроскопiчнi моделi електронних систем з орбiтальним виродженням енергетичних рiвнiв та недiагональними матричними елементами мiжелектронної взаємодiї (корельованого переносу) за допомогою конфiгурацiйного представлення. Отримано рiвняння для числового розрахунку при довiльних температурах концентрацiї полярних станiв для систем з виродженням та цiлочисельним заповненням зони $n=1$ при застосуваннi рiзних форм незбуреної густини електронних станiв. Енергетичнi спектри, отриманi в пiдходi рiвнянь руху для функцiй Ґрiна, застосованi для з’ясування ролi корельованого переносу в ефектах локалiзацiї електронiв, якi спостерiгаються в фазах Магнелi V$_{n}$O$_{2n-1}$, дихалькогенiдах перехiдних металiв NiS$_{2-x}$Se$_{x}$, фулеридах A$_{n}$C$_{60}$.
\keywords електроннi кореляцiї, енергетичний спектр, орбiтальне виродження
\end{abstract}

\end{document}